\documentstyle[a4,12pt,psfig]{article}
\def \beq{\begin{equation}}
\def \eeq{\end{equation}}
\def \beqar{\begin{eqnarray}}
\def \eeqar{\end{eqnarray}}

\begin{document}
\bigskip
\bigskip
 \rightline{}
\begin{center}
  {\bf Asymmetry distributions and mass effects 
       in dijet events at a polarized HERA 
    }\footnote{Work
    supported in part by BMBF}\\
\end{center}
\bigskip
\centerline{\rm M.~Maul ${^a}$, A.~Sch\"afer ${^a}$, 
E.~Mirkes ${^b}$,  and G.~R\"adel ${^c}$}  
\centerline
{${^a}$ \it Institute for Theoretical Physics, 93040 Regensburg, Germany.} 
\centerline
{${^b}$ \it Institute for Theoretical Particle Physics, 
76128 Karlsruhe, Germany.}                                   
\centerline
{${^c}$ \it  CERN - Div.\ PPE, 1211 Gen\`eve 23, Switzerland.}
\bigskip

\begin{quote}
The asymmetry distributions for several kinematic variables are considered 
for finding a systematic way to maximize the signal for the extraction
of the polarized gluon density. The relevance of mass effects for the
corresponding dijet cross section is discussed and the different approximations
for including mass effects are compared. We also compare via the
programs {\sc Pepsi} and {\sc Mepjet} two different Monte Carlo (MC) 
approaches for simulating the expected signal in the dijet asymmetry
at a polarized HERA.
\end{quote}
 \newpage 
\newpage
\noindent
\section{Introduction}
\begin{figure}
\psfig{figure=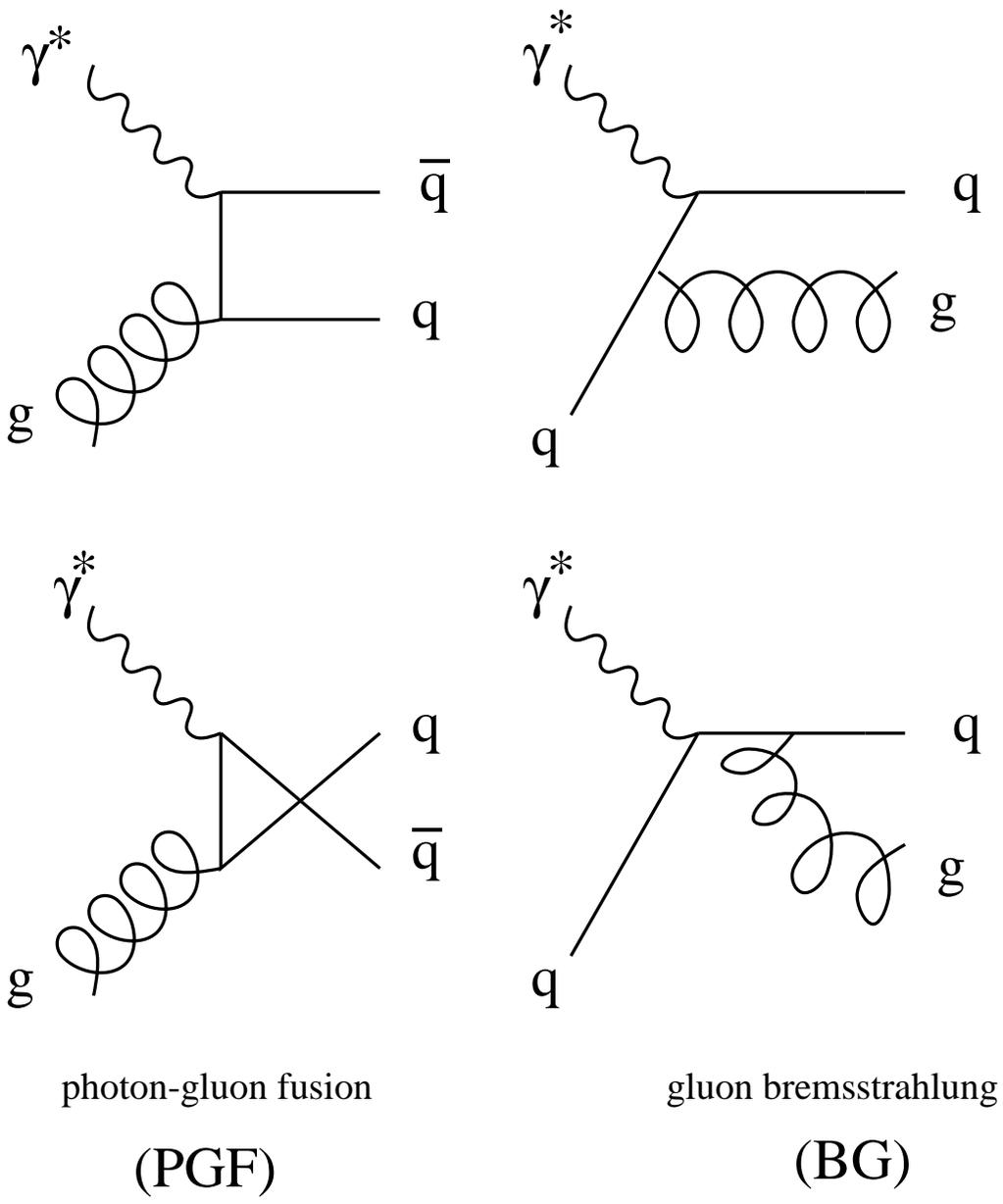,height=16cm}
\caption{Feynmangraphs of photon-gluon fusion and gluon bremsstrahlung.}
\label{fig0}   
\end{figure}

The isolation and investigation of the polarized 
gluon  parton distribution
is one of the most interesting topics for deep
inelastic scattering. The three main approaches are  
the analysis of dijet events, on which we will concentrate here, 
the $Q^2$ dependence of structure functions \cite{inclusive}, and 
the production of charmed particles \cite{semi}. 
Up-to-date parton 
distributions based on all experimental input available do not yet determine 
the form of the polarized gluon distribution with any degree of reliability
\cite{Forte}. 
This underscores the great significance of a polarized HERA, where all three
approaches described above would be feasible.
\newline
\newline
In the analysis of dijet events two different processes have to be
taken into account, the photon-gluon fusion (PGF) and the gluon 
bremsstrahlung (GB),
see Fig.~\ref{fig0}. Only the photon-gluon fusion allows to analyze the 
gluon parton distribution, while the gluon bremsstrahlung is a background
process. In the following we will investigate to which extend we
can get information about the PGF events from the polarized 
dijet cross section. For this purpose we discuss in the first section,
after some remarks to the kinematic cuts used, 
the influence of quark and remnant
masses on the dijet cross section in different MC approaches.
In the second section 
we analyze the spin asymmetry of dijet events as a function of several 
kinematic variables. We present a method of how to 
find suitable cuts in these kinematic
variables so that the significance of the asymmetry signal becomes  
largest.  This is a continuation of the discussion
started in \cite{workshop96}, where the discussion was concentrated on 
the kinematic variable $x_g$, the fractional momentum  carried by the
gluon. 
\newline
\newline
The discussion is done on the level of parton jets because on this
level the QCD improved parton model provides a well defined theory,
where comparison of different MC approaches is possible in a
quantitative way. Moreover, this will be the bases for more 
parameter-dependent additions like parton showering, 
to describe higher-order effects,
and fragmentation. We should stress, however, that the advances in
jet reconstruction justify a discussion on this level. 
For the simulations we use the following two MC generators {\sc Pepsi} and
{\sc Mepjet}. {\sc Pepsi} is a polarized add-on package to the unpolarized
 {\sc Lepto-6.5} code. The general principles
are the same as in the earlier version {\sc Pepsi~1} 
\cite{pepsi1}. The code contains massless partonic cross sections,
where the effects of heavy quark masses and remnant target masses
are taken into account by constraints reflecting energy-momentum
conservation of massive particles.
\newline
\newline
As to the {\sc Mepjet} code \cite{mepjet} we will here use two versions.
The first one neglects quark and target remnant 
masses, even charm and bottom quarks are treated as massless. 
The second one contains the quark masses in the photon-gluon fusion cross 
section, but neglects the influence of the target remnant mass.
We use this fact to demonstrate the quark and remnant mass  threshold effects 
in our simulations by comparing the {\sc Pepsi} 
with the {\sc Mepjet} results. We show that those effects are small
for QCD bremsstrahlung and large for photon-gluon fusion.  
\section{Mass effects and kinematics}
The calculations are done for HERA energies, i.e., 820 GeV protons are scattered 
off 27.5 GeV electrons. 
The cuts used are compatible with the present detectors at HERA. 
The option of polarized protons is discussed as a possible HERA 
upgrade \cite{workshop96}.
We use the following maximal kinematic cuts:
\begin{equation}
\label{eq10}
\begin{array}{lcccl}
 10^{-5} &<& x &<& 1 \\
 0.3 &<& y &<& 0.8 \\
 5~{\rm GeV}^2 &<& Q^2 &<& 100~{\rm GeV}^2 \quad . \\
\end{array}
\end{equation}
$Q^2 = -q^2$ is the virtuality of the photon, 
$x= Q^2 /(2P \cdot q)$ 
the usual Bjorken 
$x$, $y = (P \cdot q)/(P\cdot k)$, and $k$  the momentum of the
incoming lepton. The cut in $Q^2$ is guided by the idea that we look for
such a kinematic range, where the fusion process is by far dominating
the bremsstrahlung process
(see also \cite{workshop96}). This is, of course an indirect effect because
the low $Q^2$ implies small $x$, which in principle accounts for the large
gluon contribution. 
The cut $y>0.3$ does not originate from detector requirements, but is
due to the fact that the spin-dependent part of the cross section,
which is proportional to $y(1-y/2)$, becomes small
for small $y$.
The very point is that it is not possible with totally inclusive methods
to single out the photon-gluon fusion contribution. This would be possible
by triggering on charm or bottom quarks, but those particles can only 
be detected indirectly, which is
connected to losses in statistics. So, by combining both methods one can
get a quite reliable method of extracting $\Delta G$ from DIS data.
For the detection of the jets we use an algorithm which may be called a
combined $z$-$s$-cone scheme [Eq.~(\ref{eq30})]. 
This is done in order to be able to work in the same jet scheme in 
{\sc Pepsi/Lepto} and  {\sc Mepjet} because 
in this point the two codes differ substantially.
The following parameters are set:
\begin{equation}
\label{eq25}
\begin{array}{ccl}
 z_{\rm min} &=& 0.04 \\
\hat s_{\rm min} &=& 100~{\rm GeV}^2 \qquad {\rm \sc Pepsi}\\
     s_{ij, {\rm min}} &=& 100~{\rm GeV}^2 \qquad {\rm \sc Pepsi,\; Mepjet}\\
 R_{\rm min} &=& 1.0  \\
  p_{T \rm min} &=& 5~{\rm GeV} \;. \\
\end{array}   
\end{equation}
$z_{\rm min}$ is the minimum of $z = (P\cdot p_{\rm jet})/(P \cdot q)$ for the
two jet four-momenta $p_{\rm jet}$ in a dijet event. The value 
$ z_{\rm min} = 0.04$ is the default value of {\sc Lepto-6.5}.
$ \hat s_{\rm min}$ is defined
via $ \hat s := (p+q)^2$, where $p$ is the four-momentum of the incoming 
quark. The $\hat s$ cut is ineffective due to the large $p_T$ cut.
We furthermore require  for the invariant mass of the two jets
$s_{ij} := (p_1+p_2)^2> 100\; ({\rm GeV})^2 $,
where $p_1$ and $p_2$ are the two jet four-momenta. It has to be noticed that
the cuts in $Q^2$ and $p_T$ are the smallest values possible in order to
get reliable results within the LO cross sections. For smaller values
NLO effects become dominant and the LO simulations are no longer 
applicable in a reliable way. $R_{\rm min}$ defines the minimal
cone distance which two partons must have 
in order to belong to different jets. The
cone distance is given by the Pythagorean 
sum of azimuthal 
deviation $\Delta \phi$ and pseudo rapidity difference  $\Delta \eta$, i.e.:
\begin{equation}
\label{eq30}
R =\sqrt{(\Delta \eta)^2 + (\Delta \phi)^2} \quad.
\end{equation}
Finally, for $\alpha_S$ we use the one-loop 
expression with $\Lambda_{\rm QCD} = 151 \;{\rm MeV}$ (which is the
value for $\Lambda_{\rm QCD}$ over the bottom threshold), i.e.:
\begin{equation}
\label{eq40}
\alpha_S(Q^2)  = \frac{4 \pi}{(11 - \frac{2}{3} n_f) 
\ln ( Q^2 / \Lambda_{\rm QCD}^2)}  \quad.
\end{equation}
For $\alpha_{\rm em }$ we use the standard form as implemented in 
{\sc Jetset} \cite{S94}.
For the interesting asymmetries, i.e., the ratio of the polarized and
the unpolarized cross section, these coupling constants will cancel out.
Regarding the parton distributions, we use the set 
'Gluon A' from Gehrmann and Stirling \cite{GS95} as the polarized one  
and the 'GRV leading order (LO) set' given in \cite{GRV94} as the unpolarized
one.  For the observables discussed here we have checked that the qualitative 
behavior is the same for the standard LO scenario discussed in \cite{GRSV95}.
The influence of several different
models of $\Delta G $ on the polarized dijet cross section 
has been discussed in \cite{workshop96}.
\newline
\newline
%
%
We first focus on the angle $\Theta$ between the spatial momenta of the 
exchanged photon and the outgoing quark jet. From a theoretical point of
view we can define an angle $\Theta^{\rm th}$ as the angle between the
exchanged photon and the outgoing quark jet in the cm system of the two
outgoing jets. Then, in Fig.~\ref{fig1} the upward going jet 1 would be
a quark jet and the downward going jet 2 would be an anti-quark jet or
a gluon jet. Such a definition, although interesting for theoretical 
considerations, is not accessible experimentally because one cannot 
distinguish between quark, anti-quark and gluon jets. 
An experimentally accessible definition 
($\Theta^{\rm exp}$) would be that $\Theta^{\rm exp}$ is the intersection
angle (smaller than 90 degrees) between the line formed by the two jets in 
their cm system and the incoming photon.  
$\Theta^{\rm th}$ is connected with the variable $z$ from Eq.~(\ref{eq25}) via
the relation $z=\cos^2(\Theta^{\rm th}/2)$.
In order to isolate the pure spin-dependent contribution one first
regards the cross sections 
for anti-parallel ($d \sigma(\uparrow \downarrow)$) and
parallel spins ($d \sigma(\uparrow \uparrow)$) of the colliding
proton and electron beams. 
One then takes the mean and half of the difference of the differential 
cross sections of the two polarization configurations:
\begin{eqnarray}
\label{eq50}
d \sigma &=& \frac{1}{2} ( d \sigma(\uparrow \downarrow) 
                         + d \sigma(\uparrow \uparrow))
             \quad,
\nonumber \\
d \Delta \sigma &=& \frac{1}{2} ( d \sigma(\uparrow \downarrow) 
                         - d \sigma(\uparrow \uparrow))
             \quad.
\end{eqnarray}
In Fig.~\ref{fig2} we compare the polarized and unpolarized
$\Theta^{\rm th}$ distributions and the predicted asymmetries as
calculated with 
{\sc Pepsi} (filled squares) and {\sc Mepjet} (histogram).
The large deviations for the photon-gluon fusion cross sections
are due to the treatment of the quark masses and target remnant masses. 
For the shown
{\sc Mepjet} results  quark masses and target remnant masses 
are neglected completely. In
{\sc Pepsi/Lepto} they are taken into account via threshold effects 
only, i.e., the energy and momentum must be fulfilled when the
quark masses and remnant masses are added to the jet four-vectors, whereas
the cross sections in {\sc Pepsi/Lepto} are calculated for massless partons. 
As  Fig.~\ref{fig2} shows, both  effects are nearly
absent for the gluon-bremsstrahlung process, while they play 
a crucial role for the photon-gluon fusion. Moreover, the effect
due to the target remnant mass  is  dominating. Effectively
this remnant mass is included into the {\sc Lepto} code in 
such a way that not only the massive quark anti-quark pair, but
also the remnant has to be generated from the momentum 
four-vectors of the incoming photon and proton. In the case of
the gluon bremsstrahlung such an effect is absent because as
gluons are massless there is effectively no energy necessary to
radiate a gluon.
\newline
To demonstrate this we take an upgraded version of {\sc Mepjet}, 
which takes into account the exact cross section for photon
gluon fusion with quark masses and compare this to a {\sc Pepsi}
run where we substitute the target remnant by a simple massless gluon 
(see Tab.~1). In all cases we take the heavy quark masses to
be $m_{\rm c} = 1.35\; {\rm GeV}$ and $m_{\rm b} =  5 \;{\rm GeV}$, as used
in standard { \sc Jetset}.
The two cross sections are qualitatively in agreement. Note that they
may not agree exactly because {\sc Pepsi/Lepto} takes quark 
masses into account only via
threshold constraints, i.e., energy-momentum constraints inserted
into massless cross sections, while {\sc Mepjet} in the
unpolarized case takes quark masses exactly into account. Unfortunately,
there is so far no polarized version of the massive {\sc Mepjet} available. 
Besides, as the agreement shows, the {\sc Pepsi/Lepto} implementation of
quark masses  is a
quantitatively good approximation to the exact quark mass treatment, and,
moreover, the absolute effect of the quark masses in the final state
is only about 5-6\% of the total cross section. When taking into account also
the mass of the remnant, the photon-gluon fusion cross section
is reduced by another 10\%. There is no corresponding reduction in the
bremsstrahlung cross section. This is due to the fact that for emitting
a gluon, which is massless by itself, the energy can be in principle 
arbitrarily small.   

\begin{table}
\begin{center}
\begin{tabular}{|l|r|r|}
\hline &&\\
       & cross sect. [$pb$] & cross sect. [$pb$] \\
       & GB event       &  PGF-event       \\  
&&\\
\hline 
&&\\
{\sc Mepjet}(1) massless version  &   
$204.1 \pm 0.8$ & $1057.6 \pm 3.3$ \\
&& \\
{\sc Mepjet}(2) exact quark masses & 
$205.8 \pm 1.1$  & $991.5 \pm 2.5 $\\
&& \\
{\sc Pepsi/Lepto}(1)  approx. quark masses & 
$206.4 \pm 3.0 $ & $980.4 \pm 13.8 $ \\
&& \\
{\sc Pepsi/Lepto}(2) approx. quark &&\\ 
and remnant masses  & 
$205.5 \pm 3.0$ & $897.0 \pm 12.7$ \\
&&\\
\hline
\end{tabular}
\label{tab2}
\caption{Treatment of quark masses in {\sc Pepsi/Lepto} and {\sc Mepjet}
in the unpolarized case.}
\end{center}
\end{table} 
Moreover, the quark 
mass effects, as implemented in  {\sc Pepsi} enter the polarized 
and  the unpolarized cross sections in the same way. For
the asymmetries, displayed in the last
row of Fig.~\ref{fig2}, the agreement between the two programs 
is consequently nearly exact. This
on the other hand means that for asymmetry calculations we can use 
the massless program to get proper results. The numbers for 
the total cross sections are given in Tab.~1 for the unpolarized case
and in Tab.~2 for the polarized case. The errors are the
statistical MC errors, an inclusion of the systematic errors from 
parton distribution uncertainties etc. cannot be done in a reliable way.
But taking the precision of present data as used in \cite{GS95} for example,
one could accept a systematic error of at least 10\% to be realistic.
\section{Asymmetry distributions and optimization of kinematic cuts}
\begin{figure}
\psfig{figure=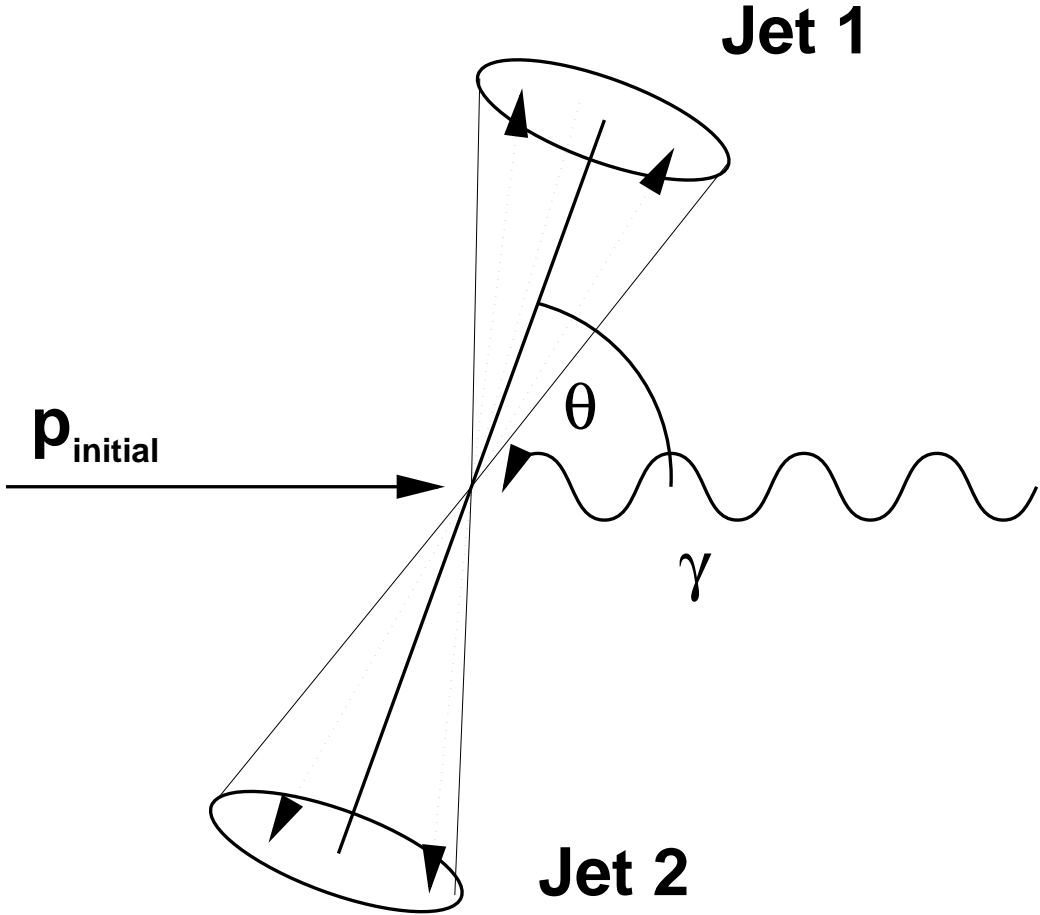,height=12cm}
\caption{Definition of the angle $\Theta$: From a theoretical point of view
an angle $\Theta^{\rm th}$ can be defined by the angle between the 
spatial momenta of 
the exchanged photon and the outgoing quark in the cm system of the two 
outgoing jets. Then the upward going jet~1 is a quark jet and the downward 
going an anti-quark or a gluon jet. But such a definition is not accessible 
experimentally because one cannot distinguish between quark, anti-quark and 
gluon jets. An experimental accessible definition ($\Theta^{\rm exp}$) is that 
$\Theta^{\rm exp}$ is the intersection angle (smaller than 90 degrees) 
between the line formed by the two jets in their cm system and the 
incoming photon line.}
\label{fig1}   
\end{figure}

\begin{figure}
\psfig{figure=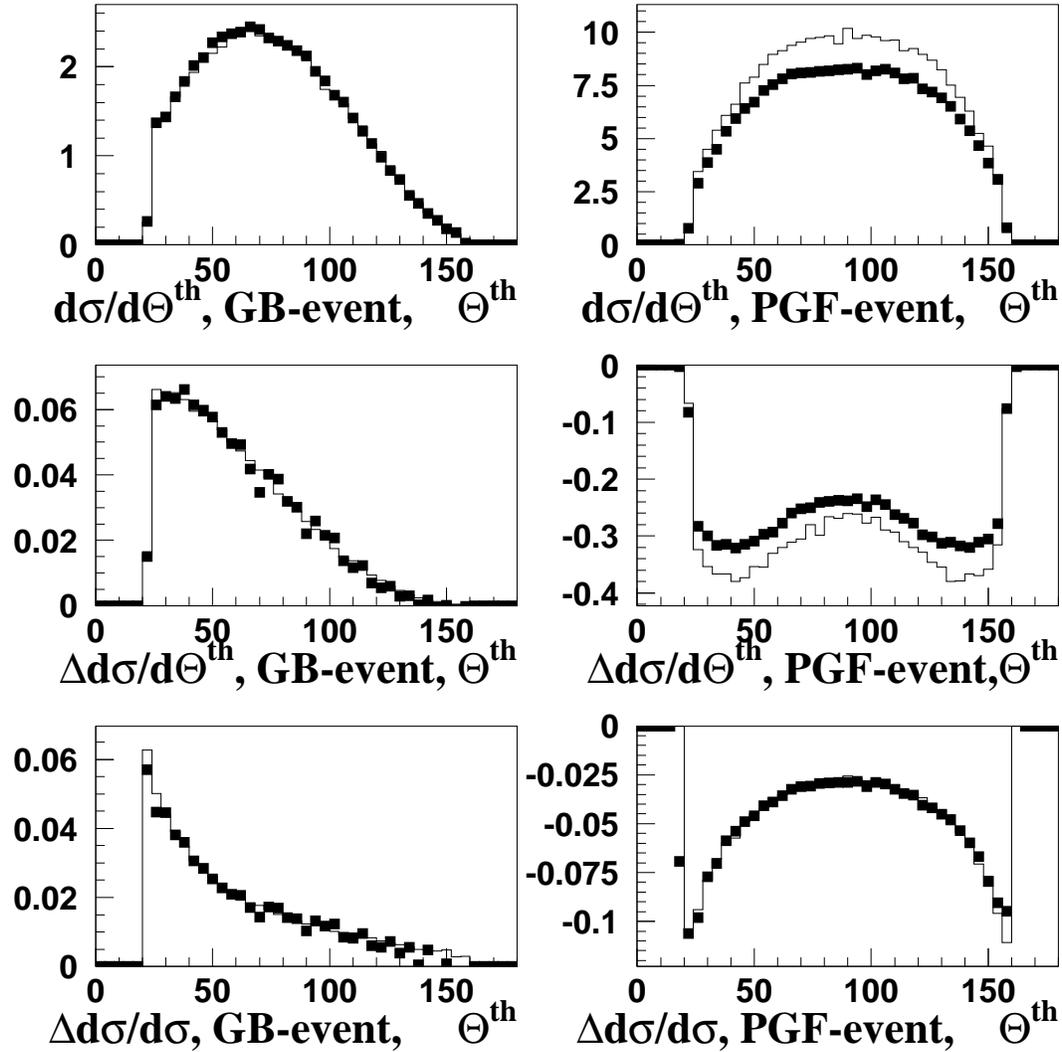,height=16cm}
\caption{Influence of quark-mass effects to the unpolarized and
polarized dijet cross sections. We show the results of the massless
{\sc Mepjet} program (histogram)
and the {\sc Pepsi} code (full squares),
where quark-mass effects and 
remnant-mass effects are included. The units are in pb/degree.}
\label{fig2}
\end{figure}

\begin{figure}
\psfig{figure=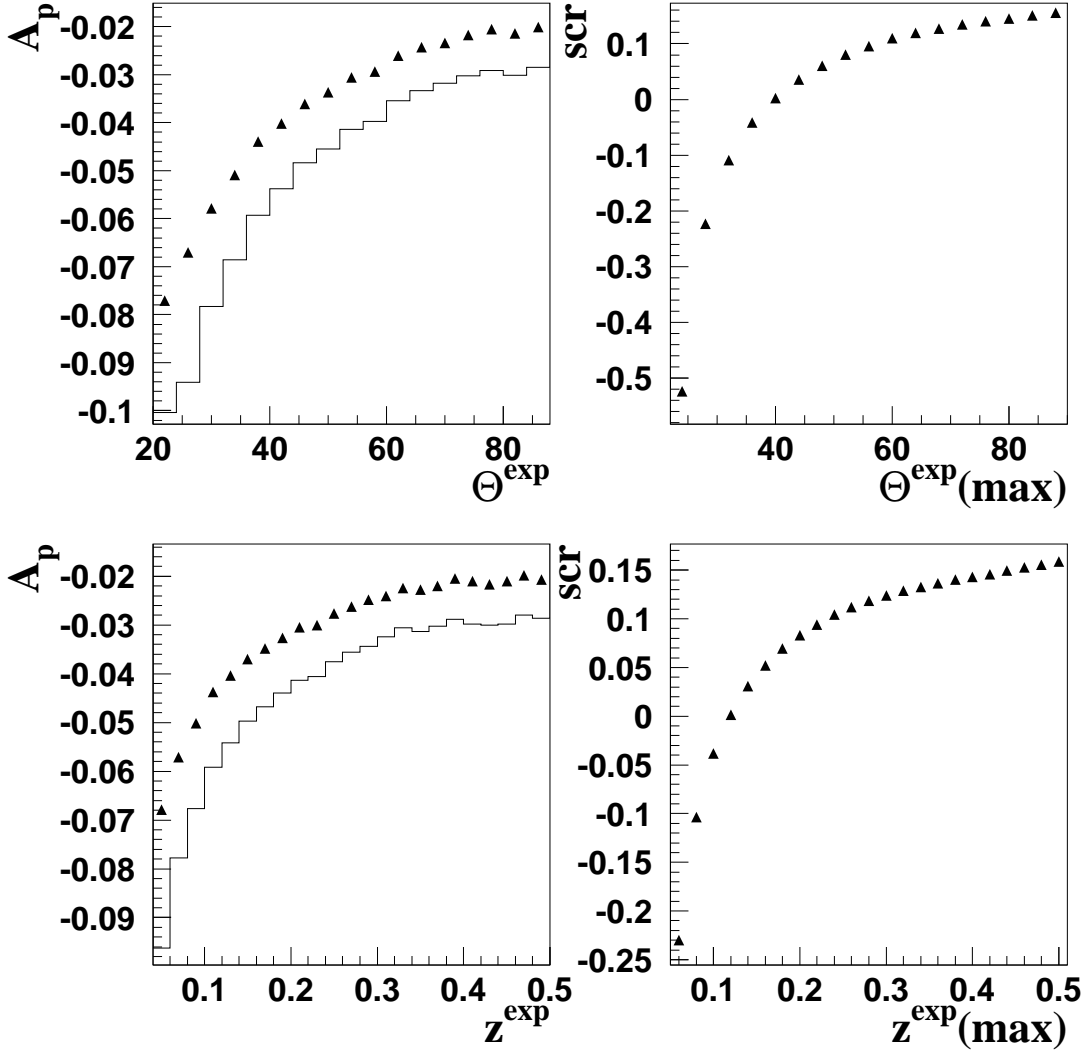,height=16cm}
\caption{Asymmetry for proton beam $A_p$ and reduced significance ($scr$) 
for the $\Theta^{\rm exp}$- and $z^{\rm exp}$ distributions. The 
histogram on the left hand side shows the asymmetry for photon-gluon 
fusion only. The triangles show the real asymmetry, which contains gluon
bremsstrahlung as well as photon-gluon fusion. The pictures on the right show 
the significance ($scr$) plotted against the  $\Theta^{\rm exp}_{\rm max}$ cut 
and $z^{\rm exp}_{\rm max}$ cut, respectively. The data are simulated with 
{\sc Pepsi}.}
\label{fig3}
\end{figure}

\begin{figure}
\psfig{figure=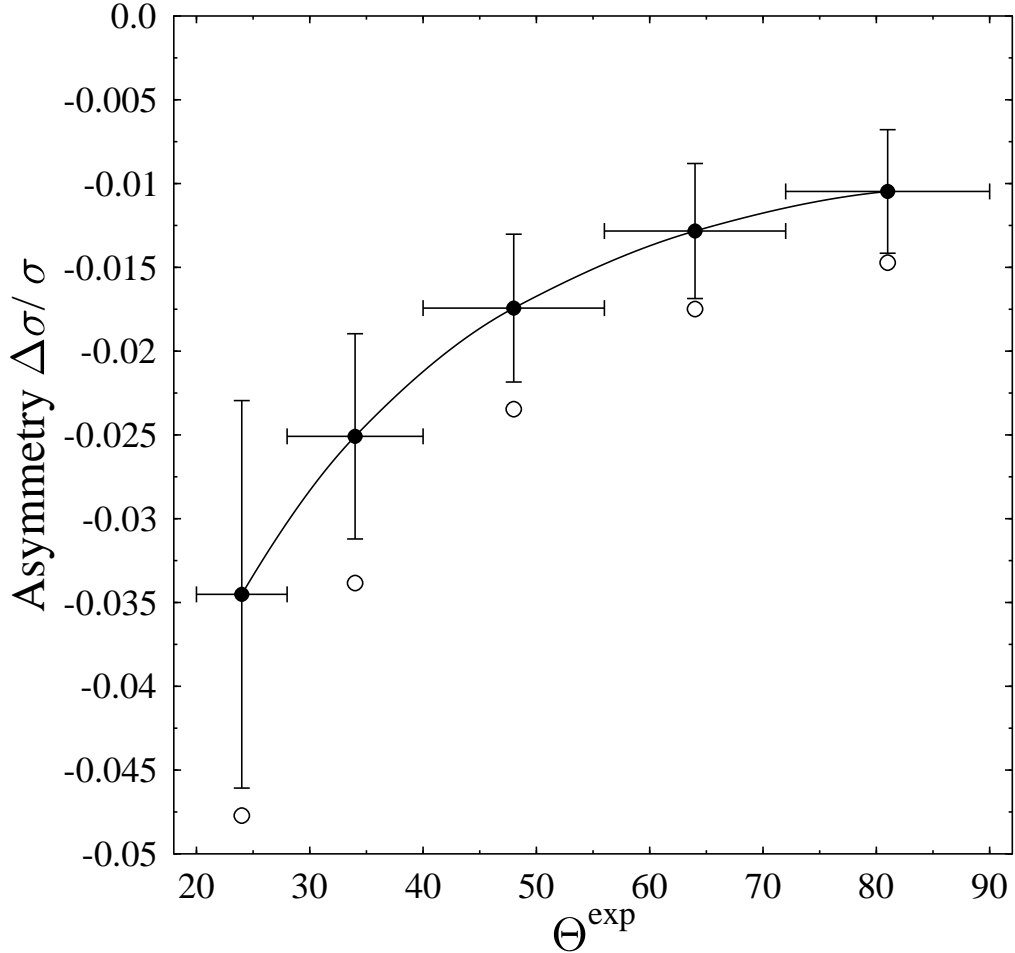,height=16cm}
\caption{Experimentally accessible asymmetry versus $\Theta^{\rm exp}$.
 The data are taken from the {\sc Pepsi} run. The full circles denote
 the total asymmetry, while the empty circles are the asymmetry 
 if the gluon bremsstrahlung process is switched off. The horizontal bars
 denote the binning width. In the plot we assumed that proton and
 electron beam polarizations are  70\% each. The error bars show 
 the expected experimental error for $100\;{\rm pb}^{-1}$
 per relative polarization.}
\label{fig4}
\end{figure}

\begin{figure}
\psfig{figure=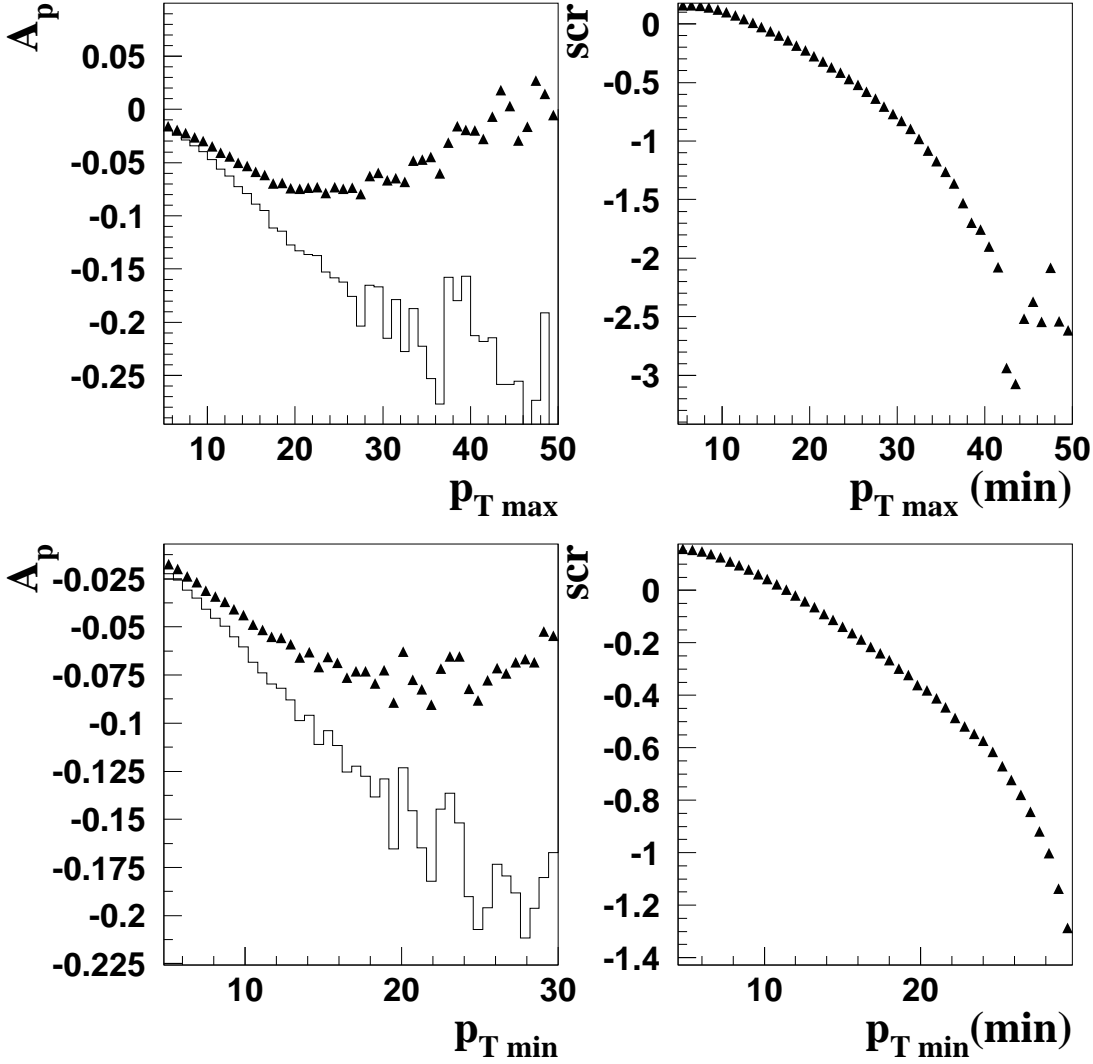,height=16cm}
\caption{Asymmetry $A_p$ and reduced significance ($scr$) for the
$p_{T,max}$- and $p_{T,min}$ distributions. The histogram on 
the left hand side shows the asymmetry for photon-gluon fusion only. 
The triangles  show the real asymmetry, where gluon
bremsstrahlung together with the photon-gluon fusion is
taken into account. The pictures on the right show the
significance versus $p_{T,max}$ and $p_{T,min}$ lower boundaries.
The data are simulated with 
{\sc Pepsi}.}
\label{fig5}
\end{figure}

\begin{figure} 
\psfig{figure=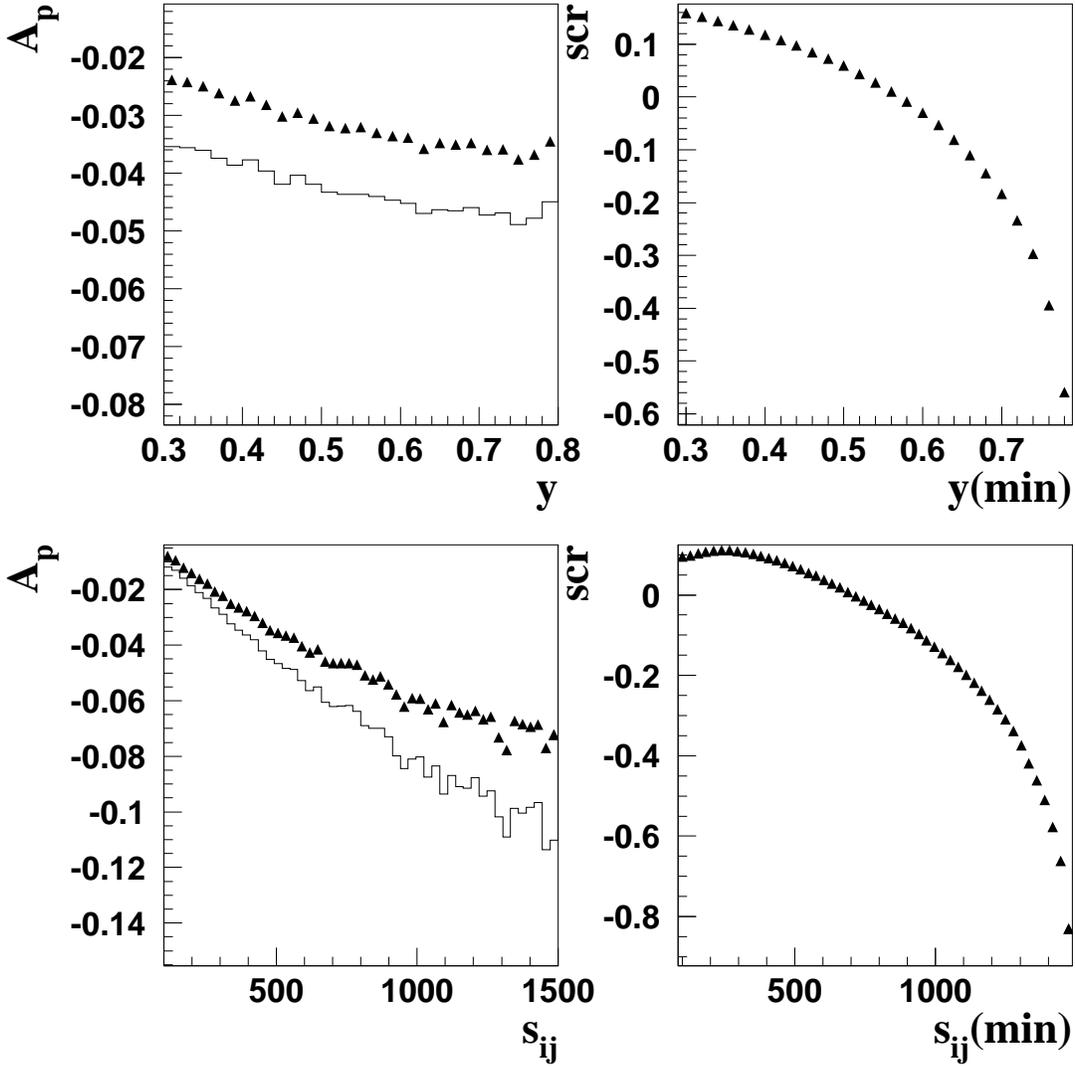,height=16cm}
\caption{Asymmetry $A_p$ and reduced significance ($scr$) for the
$y$- and $s_{ij}$ distributions. Histogram: no gluon bremsstrahlung;
triangles: full asymmetry. Data from {\sc Pepsi}.
The plots on the right show the
significance as a function of  $y$(min) and $s_{ij}$(min),
respectively.}
\label{fig6}
\end{figure}

\begin{figure}
\psfig{figure=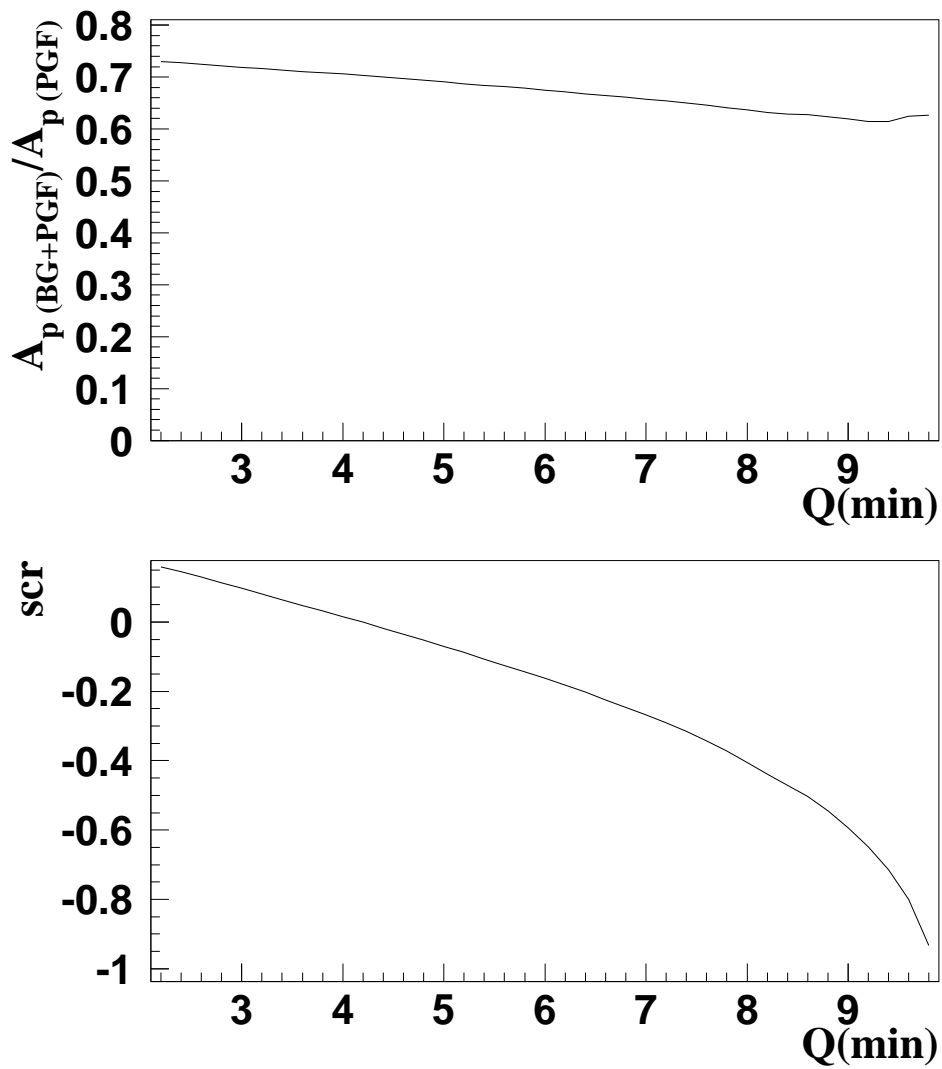,height=16cm}
\caption{Ratio of the total asymmetry to the pure gluon initiated
asymmetry (top) and reduced significance 
$scr$ as a functions of $Q$(min) (bottom);
{\sc Pepsi} data.}
\label{fig7}
\end{figure}

Fig.~\ref{fig2} shows that the asymmetry is
largest for angles $\Theta^{\rm th}$, where the axis formed by the
two jets lies close to the spatial momentum of the incoming photon, and that
the quark jets in the gluon bremsstrahlung are accumulated slightly for the 
unpolarized case and distinctly for the polarized case  close to the incoming
photon. This has quite a natural reason as the
quark jet in gluon bremsstrahlung is ordinarily more energetic
than the respective gluon jet.  
\newline
\newline
\begin{table}
\begin{center}
\begin{tabular}{|l|r|r|}
\hline &&\\
       & cross sect. [$pb$] & cross sect. [$pb$] \\
       &{\sc Pepsi}(2)       & {\sc Mepjet}(1)       \\  
&&\\
\hline 
&&\\
$\Delta$  GB-event &    3.80 $\pm$ 0.12  &   3.93 $\pm$ 0.02\\
$\Delta$ PGF-event &  -37.71 $\pm$ 0.63  &  -43.47 $\pm$ 0.15 \\
&&\\
\hline
\end{tabular}
\label{tab1}
\caption{Polarized cross sections ($\Delta \sigma$)
         for the kinematic used in the paper,
         comparing {\sc Pepsi}(2) and the  {\sc Mepjet}(1) program.}
\end{center}
\end{table} 
In Tab.~2 we give the integrated polarized cross sections. 
One should note again that the differences for the photon-gluon
fusion come from the fact that in {\sc Mepjet} quark masses are neglected.
For a detailed discussion of the influence of quark masses on
two and three jet rates see \cite{quarkmass} and references therein.
The most important question is how the significance of the asymmetry 
$A= \frac{\Delta \sigma}{\sigma}$ can be amplified by suitable 
cuts.  We will discuss several kinematic variables in the following. 
Let us first concentrate on the angle $\Theta^{\rm exp}$. As there is no
possibility to distinguish  with $100\%$ reliability quark- and
gluon jets, only the angle between the axis formed by the
dijet in its cm frame and the incoming photon in the same frame
$\Theta^{\rm exp}$ is experimentally accessible. Of course then the asymmetry 
is large for small $\Theta^{\rm exp}$, which can be already deduced from 
Fig.~\ref{fig2}, but on the other hand a cut $\Theta^{\rm exp}_{\rm max}$
also means losses in statistics. To get some feeling whether the signal
could be improved, when imposing a $\Theta^{\rm exp}_{\rm max}$ cut,
we define the {\em  significance} $sc$ by the
negative logarithm to the basis ten of the relative error in the
asymmetry $A$. (The choice of the basis ten 
is made to simplify rescaling for different luminosities.)
\beq
sc := - \log_{10} (\delta A /A) \quad.
\eeq
Moreover, the logarithmic representation has the advantage to make 
it easier to plot large numerical ranges  for $\delta A /A$.
As the asymmetry is  small for our kinematics, one can derive 
an approximative formula for
the quadratic error of the asymmetry with only the luminosity and
the unpolarized cross section:
%
%
%
%
\beq
\label{error}
\delta A = 2\sqrt{\frac{N_{\uparrow \uparrow}\cdot N_{\uparrow \downarrow}}
                       {(N_{\uparrow \uparrow}+N_{\uparrow \downarrow})^3}}
\approx  \frac{1}{\sqrt{2 L[{\rm pb}^{-1}]
 \sigma_{\rm unpol.}}} \quad.
\eeq
We can eliminate the luminosity from the significance formula by 
defining the reduced
significance $scr$:
\begin{eqnarray}
\label{significance}
sc &:=& - \log_{10} \left(\frac{2}{P A}\sqrt{
       \frac{\sigma_{\uparrow \uparrow}\cdot\sigma_{\uparrow \downarrow}}
           {(\sigma_{\uparrow \uparrow}+\sigma_{\uparrow \downarrow})^3}}
\right)
\nonumber \\
       && + \frac{1}{2}  \log_{10} (L[{\rm pb}^{-1}]) 
\nonumber \\
\nonumber \\
    &=:& scr  + \frac{1}{2} \log_{10} (P^2 L[{\rm pb}^{-1}]) \quad.
\end{eqnarray}
Here $P=P_e  P_p$ is the degree of polarization, which is the product
of the degree of polarization of the proton beam $P_p$ and the electron 
beam $P_e$, to include the effect of beam polarizations different
from 100\%. In this case the asymmetry is reduced while the absolute
error, which depends approximately only on the unpolarized cross section,
remains constant. 
In the following, if not something else is stated explicitely, the
asymmetry plots will correspond to a degree of polarization of 100\%.
The luminosity $L[{\rm pb}^{-1}]$ is given per relative 
polarization, i.e.,  $N_{\uparrow \uparrow}=
L  \sigma_{\uparrow \uparrow}$ and $N_{\uparrow \downarrow}=
L  \sigma_{\uparrow \downarrow}$.
If $sc$  is zero or even negative then the error is as large
or even larger than the signal. For a realistic experiment $sc$
should at least yield a number of 0.6, then one can set four bins 
with a two sigma  signal each. 
The definition of $scr$ allows to find the optimal cut for a maximal
significance by a look on a single diagram. It is just a very
economical way of visualizing
 the optimal parameter without trying
several MC runs.
In the following we plot the asymmetries versus several variables
together with the reduced significances $scr$ for the corresponding
minimal or maximal cuts. In order to demonstrate the usefulness of
this concept we show in Fig.~\ref{fig3} the asymmetry and $scr$
for the experimentally accessible $\Theta^{\rm exp}$ and
$z^{\rm exp}$ region. $z^{\rm exp}$ is the minimum of the two $z$ values
that belong to the two jets in Fig.~\ref{fig1}.
Both variables are closely related to each other because 
the more energetic one of the two parton jets is as compared to 
the other, i.e., the smaller $z^{\rm exp}$ is, the
closer this jet will be aligned to the incoming photon in the
frame defined in Fig.~\ref{fig1}, i.e., the
smaller is the angle 
$\Theta^{\rm exp}$ ($z^{\rm exp}=\cos^2((180^\circ-\Theta^{\rm exp})/2)$). 
The asymmetry over both variables is displayed in Fig.~\ref{fig3}. 
In both cases, the gluon bremsstrahlung reduces the asymmetry 
signal coming from photon-gluon fusion
alone because  the spin-dependent
contributions to both processes differ in sign. From the shape of the 
asymmetries (left column in Fig.~\ref{fig3}) one could suspect that a suitable
$\Theta^{\rm exp}_{\rm max}$ or $z^{\rm exp}_{\rm max}$ cut would increase 
the measurable asymmetry signal. However, the
significance plots on the right show that this is not the case. 
%
%
%
The reason for this is that the spin-dependent contribution $\Delta \sigma$ 
is small, at most ten percent of the unpolarized cross section, and therefore 
the error [see Eq.~(\ref{error})] is approximately 
anti-proportional to the square root of the unpolarized cross section alone. 
With decreasing $\Theta^{\rm exp}_{\rm max}$ or $z^{\rm exp}_{\rm max}$, 
as can be seen in Fig.~\ref{fig2}, the unpolarized cross section is reduced 
and therefore the error increases, over-compensating the gain in the 
asymmetry signal. 
As an illustration Fig.~\ref{fig4} 
shows the expected experimental errors for a luminosity 
of $L= 100\; {\rm pb}^{-1}$ per
relative polarization, i.e., $L= 200 \; {\rm pb}^{-1}$ in total. 
The assumed electron 
and proton polarization is 70\% each. The asymmetry is sizable 
using realistic cuts for the HERA machine. The envisaged luminosity 
allows to extract a clear and
distinct signal in five bins, for example. The first bin corresponds to
$\Theta^{\rm exp}_{\rm max} = 28^o$. The reduced significance for 
this cut in Fig.~\ref{fig3} is $scr =  -0.2232 $. 
For $L= 100\; {\rm pb}^{-1}$ per relative polarization and a degree of
polarization of $P=0.5$ this corresponds to a relative error of 0.334.
\newline
\newline
We now turn to the $p_T$ distributions in the lab system, 
see Fig.~\ref{fig5}, where the maximum and the minimum $p_T$ of 
the two jets is plotted. 
For pure photon-gluon fusion, the asymmetry is growing for
larger $p_{T,max}$. Taking into account gluon bremsstrahlung
the asymmetry is not only reduced, but also decreases above 
$p_{T,max}=25\; {\rm GeV}$. The behavior for $p_{T,min}$ is similar. 
Moreover, as a glance to the reduced
significances shows the relative error is increasing for increasing
$p_T$ cuts so that we recommend to choose, in spite of the
behavior of the asymmetries, the $p_T$ cut not larger than 
$5\; ({\rm GeV})^2$.  
\newline
\newline
The same is true for the variable $y$, see Fig.~\ref{fig6}, where
again by increasing the $y_{min}$ cut one gains in asymmetry but
loses significance. The only way to increase the significance
is by choosing a definite minimal cut for the invariant dijet mass
$s_{ij}$, which should be of the order $s_{ij, {\rm min}} = 250\; {\rm GeV}^2$.
However, it was shown in \cite{workshop96} that for unpolarized jet production 
it is
advisable to choose $s_{ij, {\rm min}} = 500\; {\rm GeV}^2$ in order
to minimize next to leading order effects. 
A corresponding analysis for the polarized case awaits the implementation 
of the polarized NLO cross sections into {\sc Mepjet}.
As can be seen in  
Fig.~\ref{fig6}, the losses in statistics for such cuts 
are not considerable.
\newline
\newline
The last variable discussed here is $Q$. 
The asymmetry distribution is nearly constant as a function of $Q$, so
this is a completely uninteresting variable for improving the
significance. On the other hand, as we show below,
a suitable cut in $Q$ allows to change the ratio of photon-gluon
events to gluon bremsstrahlung events. 
In Fig.~\ref{fig7} we show the ratio of the total
asymmetry and the pure gluonic asymmetry. The influence of
gluon bremsstrahlung on the asymmetry increases
for smaller  $Q_{\rm min}^2$. The significance is
 considerably larger for
small values of $Q_{\rm min}^2$, so one may think of  
choosing $Q_{\rm min}^2$ as 
small as $Q_{\rm min}^2=5\; {\rm GeV^2}$ in 
the envisaged HERA experiments for
extracting the polarized gluon density.
\newline
\newline
In summary we have found that the two programs {\sc Pepsi} and
{\sc Mepjet} (in the massless version) agree for the predicted asymmetries, 
as the mass effects enter the polarized and unpolarized cross sections 
in {\sc Pepsi} in the same way and therefore do not contribute to the 
asymmetry. For a polarized HERA the invariant
dijet mass $s_{ij, {\rm min}}$ plays a very special role, as
it is the only variable where the significance of the dijet
asymmetry signal can be slightly improved by increasing the
minimal cut. For the invariant dijet mass one should choose 
an $s_{ij, {\rm min}}$ cut between 250 and 500~GeV$^2$.
In all other cases studied here, it is advisable to choose 
small minimal cuts for $\Theta^{\rm exp},z,p_T,y,Q^2$ in order to maximize
the significance. However, the final choice of those cuts will be
dictated by future studies comparing LO and NLO polarized calculations.
We have shown that the influence of quark masses
on the gluon bremsstrahlung cross section is negligible, while it
is substantial for the photon-gluon fusion. A very dominant effect
for the photon-gluon cross section may also arise from the remnant 
mass, which makes corrections on the 10\% level.
We have furthermore  reproduced the observation
that for a polarized HERA the asymmetry signal is sizable and 
highly significant for the envisaged luminosity of $100 \; \rm pb^{-1}$
per relative polarization and 70\% beam polarization. 
\newline
\newline
{\bf Acknowledgment:} 
The authors thank G.~Ingelman, A.~De~Roeck, T.~Gehrmann, 
and H.~Ihssen for helpful discussions. M.~M. and A.~S. 
thank the BMBF, DESY and MPI f\"ur Kernphysik, Heidelberg for support.
%
\def \ajp#1#2#3{Am.~J.~Phys.~{\bf#1} (#3) #2}
\def \apny#1#2#3{Ann.~Phys.~(N.Y.) {\bf#1} (#3) #2}
\def \app#1#2#3{Acta Phys.~Polonica {\bf#1} (#3) #2 }
\def \arnps#1#2#3{Ann.~Rev.~Nucl.~Part.~Sci.~{\bf#1} (#3) #2}
\def \cmp#1#2#3{Commun.~Math.~Phys.~{\bf#1} (#3) #2}
\def \cmts#1#2#3{Comments on Nucl.~Part.~Phys.~{\bf#1} (#3) #2}
\def \cn{Collaboration}
\def \corn93{{\it lepton and Photon Interactions:  XVI International Symposium,
Ithaca, NY August 1993}, AIP Conference Proceedings No.~302, ed.~by P. Drell
and D. Rubin (AIP, New York, 1994)}
\def \cp89{{\it CP Violation,} edited by C. Jarlskog (World Scientific,
Singapore, 1989)}
\def \cpc#1#2#3{Comp.~Phys.~Commun.~{\bf#1} (#3) #2}
\def \dpff{{\it The Fermilab Meeting -- DPF 92} (7th Meeting of the American
Physical Society Division of Particles and Fields), 10--14 November 1992,
ed. by C. H. Albright \ite~(World Scientific, Singapore, 1993)}
\def \dpf94{DPF 94 Meeting, Albuquerque, NM, Aug.~2--6, 1994}
\def \efi{Enrico Fermi Institute Report No. EFI}
\def \el#1#2#3{Europhys.~Lett.~{\bf#1} (#3) #2}
\def \f79{{\it Proceedings of the 1979 International Symposium on lepton and
Photon Interactions at High Energies,} Fermilab, August 23-29, 1979, ed.~by
T. B. W. Kirk and H. D. I. Abarbanel (Fermi National Accelerator Laboratory,
Batavia, IL, 1979}
\def \hb87{{\it Proceeding of the 1987 International Symposium on lepton and
Photon Interactions at High Energies,} Hamburg, 1987, ed.~by W. Bartel
and R. R\"uckl (Nucl. Phys. B, Proc. Suppl., vol. 3) (North-Holland,
Amsterdam, 1988)}
\def \ib{{\it ibid.}~}
\def \ibj#1#2#3{~{\bf#1} (#3) #2}
\def \ichep72{{\it Proceedings of the XVI International Conference on High
Energy Physics}, Chicago and Batavia, Illinois, Sept. 6--13, 1972,
edited by J. D. Jackson, A. Roberts, and R. Donaldson (Fermilab, Batavia,
IL, 1972)}
\def \ijmpa#1#2#3{Int.~J.~Mod.~Phys.~A {\bf#1} (#3) #2}
\def \ite{{\it et al.}}
\def \jmp#1#2#3{J.~Math.~Phys.~{\bf#1} (#3) #2}
\def \jpg#1#2#3{J.~Phys.~G {\bf#1} (#3) #2}
\def \lkl87{{\it Selected Topics in Electroweak Interactions} (Proceedings of
the Second Lake Louise Institute on New Frontiers in Particle Physics, 15--21
February, 1987), edited by J. M. Cameron \ite~(World Scientific, Singapore,
1987)}
\def \ky85{{\it Proceedings of the International Symposium on lepton and
Photon Interactions at High Energy,} Kyoto, Aug.~19-24, 1985, edited by M.
Konuma and K. Takahashi (Kyoto Univ., Kyoto, 1985)}
\def \mpla#1#2#3{Mod.~Phys.~Lett.~A {\bf#1} (#3) #2}
\def \nc#1#2#3{Nuovo Cim.~{\bf#1} (#3) #2}
\def \npb#1#2#3{Nucl.~Phys.~B {\bf#1} (#3) #2}
\def \pw#1#2#3{Part.~World.~{\bf#1} (#3) #2}   
\def \pisma#1#2#3#4{Pis'ma Zh.~Eksp.~Teor.~Fiz.~{\bf#1} (#3) #2[JETP Lett.
{\bf#1} (#3) #4]}
\def \pl#1#2#3{Phys.~Lett.~{\bf#1} (#3) #2}
\def \plb#1#2#3{Phys.~Lett.~B {\bf#1} (#3) #2}
\def \pr#1#2#3{Phys.~Rev.~{\bf#1} (#3) #2}
\def \pra#1#2#3{Phys.~Rev.~A {\bf#1} (#3) #2}
\def \prd#1#2#3{Phys.~Rev.~D {\bf#1} (#3) #2}
\def \prl#1#2#3{Phys.~Rev.~Lett.~{\bf#1} (#3) #2}
\def \prp#1#2#3{Phys.~Rep.~{\bf#1} (#3) #2}
\def \ptp#1#2#3{Prog.~Theor.~Phys.~{\bf#1} (#3) #2}
\def \rmp#1#2#3{Rev.~Mod.~Phys.~{\bf#1} (#3) #2}
\def \rp#1{~~~~~\ldots\ldots{\rm rp~}{#1}~~~~~}
\def \zpc#1#2#3{Z.~Phys.~C {\bf#1} (#3) #2}

\newpage

\begin{thebibliography}{99}

\bibitem{inclusive}
T.~Gehrmann, W.~J.~Stirling \prd{53}{1996}{6100}; \\
M.~Gl\"uck, E.~Reya, M.~Stratmann, W.~Vogelsang,
\prd{53}{1996}{4775}.



\bibitem{semi}
A.~D.~Watson, \zpc{12}{1982}{123}; \\
G.~Altarelli, W.~J.~Stirling, \pw{1}{1989}{40};\\
M.~ Gl\"uck, E.~Reya, W.~Vogelsang \npb{351}{1991}{579};\\
T.~Morii, S.~Tanaka, T.~Yamanishi \plb{322}{1994}{253};\\ 
N.~I.~Kochelev, T.~Morii, T.~Yamanishi, \plb{405}{1997}{168}. 


\bibitem{Forte}
S.~Forte  {\em Polarized structure functions: a status report}.
DFTT-57-96, Oct 1996. 15pp. Invited talk at 14th International Conference on
Particles and Nuclei (PANIC 96), Williamsburg, VA, 22-28 May 1996, and at 12th International Symposium on
High-energy Spin Physics (SPIN96), Amsterdam, Netherlands, 10-14 Sep 1996. 
e-Print Archive: hep-ph/9610238. 


\bibitem{workshop96}
A.~De Roeck, J.~Feltesse, F.~Kunne, M.~Maul, A.~Sch\"afer, C.Y.~Wu,
E.~Mirkes, G.~R\"adel,
{\em prospects for measuring $\Delta G$ from jets at HERA with polarized 
protons and electrons}
In *Hamburg 1995/96, Future physics at HERA* 803-814. 
e-Print Archive: hep-ph@xxx.lanl.gov - 9610315;\\
J.~Feltesse, F.~Kunne, E.~Mirkes,
Phys.~Lett.~{\bf B388} (1996) 832. 


\bibitem{lepto65}
G.~Ingelman, A.~Edin, J.~Rathsman, 
\cpc{101}{1997}{108}.

\bibitem{pepsi1}
L.~Mankiewicz, A.~Sch\"afer, M.~Veltri, \cpc{71}{1992}{305}. 


\bibitem{mepjet}
E.~Mirkes, D.~Zeppenfeld ,
Phys. Lett. {\bf B380} (1996) 205;
The program can be obtained on request by the authors.


\bibitem{quarkmass}
G.~Vicente and R.~Garcia, Oct. 1996. 95pp. Ph.D. Thesis. 
e-Print Archive: hep-ph@xxx.lanl.gov - 9703359.


\bibitem{S94}
T.~Sj\"ostrand, Computer Pysics Commun. {\bf 82} (1994) 74.

\bibitem{GS95}
T.~Gehrmann, W.~J.~Stirling,  \prd{53}{1996}{6100}.

\bibitem{GRSV95}
M.~Gl\"uck, E.~Reya, M.~Stratmann, W.~Vogelsang,
\prd{53}{1996}{4475}.



\bibitem{GRV94}
M.~Gl\"uck, E.~Reya, A.~Vogt, \zpc{67}{1995}{433}. 

\end{thebibliography}
\end{document}